\documentstyle[amssymb,rmp,aps]{revtex}

\begin{document}
\title{Metal-Insulator Transition of the Quasi-One Dimensional Luttinger Liquid Due
to the Long-Range Character of the Coulomb Interaction}
\author{V. S. Babichenko}
\address{RSC ''Kurchatov Institute'', Moscow 123182, Russia. \\
e-mail: babichen@kurm.polyn.kiae.su}
\date{\today}
\maketitle

\begin{abstract}
An instability of the quasi-1D Luttinger liquid associated with the metal -
insulator transition is considered. It is shown that the homogeneous metal
ground state of this liquid is unstable and the charge density wave arises
in the system. The wave vector of this charge density wave has nonzero
component both along the direction of the chains and in the perpendicular
direction. \ The ground state has a dielectric gap at the Fermi surface, the
value of this gap being calculated.
\end{abstract}

1. The instability of a quasi-1D electron liquid with respect to the
metal-insulator transition with the formation of the charge density wave
(CDW) is analyzed in the present work. The materials, which can be
considered as candidates for the theory proposed below are quasi-1D organic
conductors, for example, (TMTSF)$_{2}$X in strong magnetic fields \ \ \ [1,
2].

The motion of particles is supposed to take place along the 1D semiconductor
chains which form the crystal lattice with the lattice constant $\lambda $
and represent a quasi-1D doped semiconductor. The thickness of the chains $a$
is supposed to be much smaller than the distance between chains $\lambda $ \
\ \ ($a<<\lambda $ ). The interaction between particles is the usual 3D
Coulomb interaction $V_{\vec{r},\vec{r}^{\prime }}=e^{\ast 2}/\mid \vec{r}-%
\vec{r}^{\prime }\mid $ with the static dielectric constant $\epsilon _{0}$
. So, the effective charge of particles is $e^{\ast 2}=e^{2}/$ $\epsilon _{0}
$ where $e$ is the bare electron charge. The analysis of the role of the
long-range Coulomb interaction character is the main goal of the work.

For the first time the different types of instabilities of a quasi-1D Fermi
liquid with the short-range bare interaction in the system of the ordered
metallic chains in the parquet approximation have been investigated in the
work [3]. The quasi-1D Luttinger liquid with the long-range Coulomb
interaction has been investigated in the work [4]. In that work only the
forward scattering of particles has been taken into account. The influence
of the long-range character of the Coulomb interaction in carbon nanotube
which can be considered as consisting of two chains has been analyzed in
[5]. In this case the Coulomb interaction between two chains does not change
the Luttinger character of the 1D electron liquid. The metal-insulator
transition of the electron liquid in a strong magnetic field in the parquet
approximation in the case when the Coulomb interaction can be considered as
the short-range interaction has been considered in the work [6]. Note that
the long-range character of the Coulomb interaction can change the situation
from the parquet, when several channels have to be taken into account [10],
to the situation with only one separated channel [9]. The long-range tail of
the Coulomb interaction in the electron liquid in a strong magnetic field
has been considered in [7]. In the present work the influence of the
long-range character of the \ Coulomb interaction \ on the properties of the
backward scattering in the quasi-1D electron liquid is analyzed. In
addition, the influence of the short-range correlations is taken into
account too. If the Coulomb interaction is supposed to have only the short
range character, i.e. the electrons interact only at the same chain, the
ground state has the properties of the Luttinger liquid. The influence of
the long-range part of the Coulomb interaction, i.e. when the interaction
between electrons situated at the different chains is essential, results in
the instability of the metallic homogeneous ground state of the Luttinger
liquid. In addition, the ground state becomes dielectric and inhomogeneous.
The charge-density wave (CDW) arises in the system. This CDW has the
wavevector components both along the direction of chains and in the
perpendicular direction. The wavevector component along the direction of
chains equals $k_{z}=Q=2p_{F}/\hbar $ where $p_{F\text{ }}$ is the Fermi
momentum $p_{F\text{ }}=\hbar \pi n\lambda ^{2}$ \ and $n$ is the density of
the Fermi liquid. The wave vector component perpendicular to the direction
of chains equals $\mid \overrightarrow{K}_{\perp }\mid =\sqrt{2}\pi /\lambda 
$. In this case the amplitude of the CDW does not depend on the chain
coordinate, but the phase changes its value from zero to $\pi $ for the
neighboring chains. The ground state of the system is dielectric and has a
dielectric gap at the Fermi surface. Below we use the system of units with $%
\hbar =1$. \ 

The essential point of our analysis is the consideration of the interference
between the channel of the backward scattering and the channel, which
results in the infrared divergences, namely, the channel of the forward
scattering.

The transitions between different chains are neglected, but they do not play
an essential role if they are small in comparison with the dielectric gap
value arising at the Fermi surface. We suppose that each chain can be
considered as a continuous medium and the system of chains forms the square
lattice with the lattice constant $\lambda $. The parameters of the system
is assumed to obey the inequalities $\frac{1}{a_{B}}\ll p_{F}\ll \frac{1}{%
\lambda }$ where $\ a_{B}=1/m^{\ast }e^{\ast 2}$ \ is the effective Bohr
radius and $m^{\ast }$ is the electron effective mass. The first part of
this inequality means that the non-dimensional constant $\alpha =e^{\ast
2}/v_{F}$ , \ which defines the value of the effective interaction, is small 
$\alpha <<1$ where $v_{F}$ is the Fermi velocity. Because of the condition $%
\alpha <<1$ the random phase approximation (RPA) is correct for the
calculation of the thermodynamic properties, for example, free energy [11].
The second part of the inequality means that the density of the electron
liquid is sufficiently small so that the Coulomb correlations are large and
the long-range character of the Coulomb interaction is pronounced. Below,
for the simplicity of notations, the effective charge $e^{\ast }$ will be
denoted by $e$.

The electron liquid is supposed to be spinless. This situation can be
realized by the switching-on the strong magnetic field directed along the
chains. In this case the electron spins are frozen, but the motion of the
electrons along the chains are not perturbed by the magnetic field.
Moreover, the jumps between chains are suppressed by the magnetic field. 

2. The action of a quasi-1D Fermi liquid with the Coulomb interaction has
the form

\begin{equation}
S=S_{0}+S_{int}  \label{1}
\end{equation}

\[
S_{0}=\sum\limits_{\overrightarrow{R}}\int dzdt\bar{\psi}(t,\vec{r})\left(
i\partial _{t}+\mu -\hat{p}_{z}^{2}/2m\right) \psi (t,\vec{r}) 
\]

\[
S_{int}=-\frac{1}{2}\sum\limits_{\overrightarrow{R},\overrightarrow{R}%
^{\prime }}\int dtdzdz^{\prime }\left( \bar{\psi}(t,\vec{r})\psi (t,\vec{r}%
)\right) V_{\vec{r},\vec{r}^{\prime }}\left( \bar{\psi}(t,\vec{r}^{\prime
})\psi (t,\vec{r}^{\prime })\right) 
\]
where $z$ -is the direction of chains, $\widehat{p}_{z}=-i\partial _{z}$ , $%
\vec{r}=$ $\left( z,\overrightarrow{R}\right) $,$\ $ $\overrightarrow{R}$
are the discrete coordinates \ of a chain in the plane perpendicular to the
direction of chains. \ The field $\psi (t,\vec{r})$ is the Fermi field
(Grassmann variables). For simplicity, we suppose that the lattice is the
square lattice with the lattice constant $\lambda $. The motion of electrons
from one chain to another is neglected.

The correlation properties of the Fermi liquid are connected with the
correlations of quasi-particles near the Fermi surface. The fields of these
quasi-particles are denoted by $\psi ^{\left( +\right) }$ for the part of
the Fermi surface with $p_{z}=+p_{F}$ and $\psi ^{\left( -\right) }$ for $%
p_{z}=-p_{F}$ .

The Coulomb interaction can be decoupled by the introduction of the virtual
plasmon fields of $\phi $ and $\Phi $ [7]. The plasmon field $\phi $ has the
small z-component of the momentum transfer $k_{z}$\ and corresponds to the
forward scattering of quasi-particles and the plasmon field $\Phi $\ has the
z-component of the momentum transfer $k_{z}$\ close to the value $Q=2p_{F}$
and corresponds to the backward scattering. The action of quasi-particles
can be represented in the form

\begin{equation}
S_{L}[\Psi ,\phi ,\Phi ]=S_{F}[\Psi ,\phi ,\Phi ]+S_{forward}^{\left(
0\right) }\left[ \phi \right] +S_{back}^{\left( 0\right) }\left[ \Phi \right]
,  \label{2}
\end{equation}
where

\begin{equation}
S_{F}=\int\limits_{0}^{\beta }dt\int dz\sum\limits_{\overrightarrow{R}%
}\left( 
\begin{array}{cc}
\bar{\psi}^{\left( +\right) }; & \bar{\psi}^{\left( -\right) }
\end{array}
\right) \left( 
\begin{array}{cc}
\left( G_{\phi }^{\left( +\right) }\right) ^{-1} & -i\Phi \\ 
-i\Phi ^{+} & \left( G_{\phi }^{\left( -\right) }\right) ^{-1}
\end{array}
\right) \left( 
\begin{array}{c}
\psi ^{\left( +\right) } \\ 
\psi ^{\left( -\right) }
\end{array}
\right)  \label{SFermi}
\end{equation}

\begin{equation}
S_{forward}^{\left( 0\right) }\left[ \phi \right] =-\frac{1}{2}%
\int\limits_{0}^{\beta }dt\int dk_{z}d^{2}\vec{k}_{\perp }\phi (t,-k_{z},-%
\vec{k}_{\perp })V^{-1}\left( k_{z},\vec{k}_{\perp }\right) \phi (t,k_{z},%
\vec{k}_{\perp })  \label{SForward0}
\end{equation}

\begin{equation}
S_{back}^{\left( 0\right) }\left[ \Phi \right] =-\int\limits_{0}^{\beta
}dt\int dk_{z}d^{2}\vec{k}_{\perp }\Phi ^{+}(t,k_{z},\vec{k}_{\perp
})U^{-1}\left( \vec{k}_{\perp }\right) \Phi (t,k_{z},\vec{k}_{\perp })
\label{SBack0}
\end{equation}
In these expressions the momentum $\overrightarrow{k}_{\perp }$ belongs to
the elementary cell of the reciprocal lattice formed by the system of
chains, $\overrightarrow{k}_{\perp }$ is perpendicular to the chain
direction z and $\beta $ is the inverse temperature. The temperature is
supposed equal to zero, but the Matsubara technique is used. The functions $%
G_{\phi }^{\left( \pm \right) }$ are the Green functions of quasi-particles
near the different parts of the Fermi surface with $p_{z}=\pm p_{F}$ in the
external plasmon field $\phi $ , $\widehat{p}_{z}=-i\partial _{z}$ \ and $%
\overrightarrow{R}$ are the discrete coordinates \ of chains in the plane
perpendicular to the chain direction.

\begin{equation}
\left( G_{\phi }^{\left( \pm \right) }\right) ^{-1}=-\partial _{\tau }\mp
v_{F}\widehat{p}_{z}-i\phi (t,z,\overrightarrow{R}),  \label{6}
\end{equation}
The function $V\left( k_{z},\vec{k}_{\perp }\right) $ is the Fourier
transformations of the Coulomb potential with the small \ momentum transfer $%
k_{z}$ , and $U\left( \vec{k}_{\perp }\right) $ is the Coulomb potential
with the component of the momentum transfer $k_{z}$ close to the value of $%
k_{z}=Q=2p_{F}$

\begin{equation}
V\left( k_{z},\vec{k}_{\perp }\right) =\sum\limits_{\overrightarrow{b}}\frac{%
4\pi e^{2}}{k_{z}^{2}+\left( \overrightarrow{k}_{\perp }+\overrightarrow{b}%
\right) ^{2}}\text{ \ ; \ \ }U\left( \vec{k}_{\perp }\right) =\sum\limits_{%
\overrightarrow{b}}\frac{4\pi e^{2}}{Q^{2}+\left( \overrightarrow{k}_{\perp
}+\overrightarrow{b}\right) ^{2}}  \label{7}
\end{equation}
Here $\overrightarrow{b}$ are the vectors of the reciprocal lattice so that $%
\overrightarrow{b}=\left( 2\pi n/\lambda ;2\pi m/\lambda \right) $ where $%
n,m $ are the integer numbers, and $\vec{k}_{\perp }$ belongs to the cell of
the reciprocal lattice $\left( -\pi /\lambda <k_{x,y}<\pi /\lambda \right) $
. It is convenient to extract the term with $\overrightarrow{b}=0$ in the
sums (\ref{7}) . As a result, the sums (\ref{7}) can be represented in the
form of sums of two items. The first one is the summand with $%
\overrightarrow{b}=0 $ and the second is a sum over $\overrightarrow{b}\neq 0
$ in (\ref{7}). Thus $U\left( \vec{k}_{\perp }\right) $ and $V\left( k_{z},%
\vec{k}_{\perp }\right) $ are represented in the form of the
pseudo-potentials having long-range and short-range parts

\begin{equation}
V\left( k_{z},\vec{k}_{\perp }\right) =\frac{4\pi e^{2}}{k_{z}^{2}+%
\overrightarrow{k}_{\perp }^{2}}+V_{core}\left( \vec{k}_{\perp }\right)
\label{VPseudoPot}
\end{equation}

\begin{equation}
U\left( \vec{k}_{\perp }\right) =\frac{4\pi e^{2}}{Q^{2}+\overrightarrow{k}%
_{\perp }^{2}}+U_{core}\left( \vec{k}_{\perp }\right)  \label{UPseudoPot}
\end{equation}
The functions $V_{core}\left( \vec{k}_{\perp }\right) $ and $U_{core}\left( 
\vec{k}_{\perp }\right) $ can be considered as independent of $\vec{k}%
_{\perp }$ if $\mid \vec{k}_{\perp }\mid <<\Lambda $. In the case of the
small thickness of chains $a<<\lambda $, which is supposed, the dependence
of $V_{core}$ and $U_{core}$ on the momentum $\vec{k}_{\perp }$ is weak for $%
\mid \vec{k}_{\perp }\mid \thicksim \Lambda $ \ too and thus can be neglected

\begin{equation}
V_{core}\left( \vec{k}_{\perp }\right) \thickapprox U_{core}\left( \vec{k}%
_{\perp }\right) \thickapprox V_{core}\left( 0\right) =U_{core}\left(
0\right) =\frac{4\pi e^{2}}{\widetilde{\Lambda }^{2}}  \label{VUcore}
\end{equation}
The value $\widetilde{\Lambda }$ in (\ref{VUcore}) has the form

\[
\widetilde{\Lambda }^{2}=(2/\pi )\Lambda ^{2}\ln ^{-1}\left( \frac{\Lambda
_{\infty }}{\Lambda }\right) , 
\]
where $\Lambda =\pi /\lambda $, $\Lambda _{\infty }\thicksim 1/a$, thus $%
\Lambda _{\infty }>>\Lambda $ and $\widetilde{\Lambda }<<\Lambda $. The
parameter $\Lambda _{\infty }$ is the cutoff momentum in the sum over $%
\overrightarrow{b}$ in (\ref{7}). The truncation of the sum (\ref{7})\ is
the necessary operation in the connection with the divergency of this sum at
the large momentums $\overrightarrow{b}$. The Fermi momentum is supposed to
be sufficiently small, so the inequality $Q<<\widetilde{\Lambda }<<$ $%
\Lambda $\ obeys. The functions $V\left( k_{z},\vec{k}_{\perp }\right) $ and 
$U\left( \vec{k}_{\perp }\right) $ (\ref{7}) \ as well as $V^{-1}\left(
k_{z},\vec{k}_{\perp }\right) $ and $U^{-1}\left( \vec{k}_{\perp }\right) $
are the periodic functions of $\vec{k}_{\perp }$ with the periods $%
\overrightarrow{b}_{x,y}=\left( 2\pi /\lambda ;0\right) ;\left( 0;2\pi
/\lambda \right) $ and $\vec{k}_{\perp }$ in\ the expressions (\ref
{VPseudoPot}), (\ref{UPseudoPot})\ belongs to the elementary cell of the
reciprocal lattice. The function $U^{-1}\left( \vec{k}_{\perp }\right) $ can
be represented in the form

\begin{equation}
U^{-1}\left( \vec{k}_{\perp }\right) =U_{core}^{-1}\left( \vec{k}_{\perp
}\right) \left[ 1-\frac{4\pi e^{2}U_{core}^{-1}\left( \vec{k}_{\perp
}\right) }{\overrightarrow{k}_{\perp }^{2}+Q^{2}+4\pi
e^{2}U_{core}^{-1}\left( \vec{k}_{\perp }\right) }\right]  \label{Uinv(k)}
\end{equation}
\ The coordinate representation of $U^{-1}\left( \vec{k}_{\perp }\right) $
can be written as 
\begin{equation}
\left( U^{-1}\right) _{\overrightarrow{R}_{1},\overrightarrow{R}%
_{2}}=U_{core}^{-1}\left( \overrightarrow{R}_{1}-\overrightarrow{R}%
_{2}\right) -4\pi e^{2}\left( U_{core}^{-1}\left( 0\right) \right)
^{2}C\left( \overrightarrow{R}_{1}-\overrightarrow{R}_{2}\right)
\label{InvCoul}
\end{equation}
where $C\left( \overrightarrow{R}\right) $ is

\begin{equation}
C\left( \overrightarrow{R}\right) =\int \frac{d^{2}k_{\perp }}{\left( 2\pi
\right) ^{2}}\frac{\exp \left( i\vec{k}_{\perp }\overrightarrow{R}\right) }{%
\overrightarrow{k}_{\perp }^{2}+Q^{2}+4\pi e^{2}U_{core}^{-1}\left( \vec{k}%
_{\perp }\right) }  \label{C(R)}
\end{equation}
The region of integration in (\ref{C(R)}) is the elementary cell of the
reciprocal lattice. The value $4\pi e^{2}U_{core}^{-1}\left( \vec{k}_{\perp
}\right) $ satisfies the estimations $4\pi e^{2}U_{core}^{-1}\left( \vec{k}%
_{\perp }\right) \thickapprox 4\pi e^{2}U_{core}^{-1}\left( 0\right) =%
\widetilde{\Lambda }^{2}<<\Lambda ^{2}$\ . Thus, the momentums $\vec{k}%
_{\perp }$, which give the main contribution to the integral (\ref{C(R)})
for the function $C\left( \overrightarrow{R}\right) $, are of the order of $%
\widetilde{\Lambda }<<\Lambda $. For this reason, the function $%
U_{core}^{-1}\left( \vec{k}_{\perp }\right) $ in the equality (\ref{C(R)})
can be taken equal to $U_{core}^{-1}\left( 0\right) $\ . Thus, we obtain 
\begin{eqnarray}
C\left( \overrightarrow{R}\right) &=&\frac{1}{2\pi }K_{0}\left( \mid 
\overrightarrow{R}\mid \widetilde{\Lambda }_{Q}\right) \text{ \ \ \ \ \ \ \
\ \ \ \ \ \ \ \ \ \ \ \ \ \ \ \ \ \ \ \ \ \ \ \ \ \ \ for }\mid 
\overrightarrow{R}\mid \Lambda >>1 \\
C\left( \overrightarrow{R}\right) &=&\frac{1}{2\pi }\ln \left( \frac{\Lambda 
}{\widetilde{\Lambda }_{Q}}\right) \text{ }=\frac{1}{2\pi }\text{ }%
K_{0}\left( \frac{\widetilde{\Lambda }_{Q}}{\Lambda }\right) \text{\ \ \ \ \
\ \ \ \ \ \ \ for }\mid \overrightarrow{R}\mid \Lambda \lesssim 1  \nonumber
\end{eqnarray}
where $K_{0}\left( x\right) $ \ is the cylinder function of imaginary
argument and $\widetilde{\Lambda }_{Q}=\sqrt{\widetilde{\Lambda }^{2}+Q^{2}}$%
\ . The value of $\mid \overrightarrow{R}_{1}-\overrightarrow{R}_{2}\mid $
in the equality (\ref{InvCoul}) is supposed to obey the inequality $\mid 
\overrightarrow{R}_{1}-\overrightarrow{R}_{2}\mid >>1/\Lambda $. If $\mid 
\overrightarrow{R}_{1}-\overrightarrow{R}_{2}\mid >>1/\widetilde{\Lambda }%
_{Q}\simeq 1/\widetilde{\Lambda }$ , the function $K_{0}\left( \mid 
\overrightarrow{R}_{1}-\overrightarrow{R}_{2}\mid \widetilde{\Lambda }%
_{Q}\right) $ can be replaced by

\[
K_{0}\left( \mid \overrightarrow{R}_{1}-\overrightarrow{R}_{2}\mid 
\widetilde{\Lambda }_{Q}\right) \rightarrow \frac{2\pi }{\widetilde{\Lambda }%
^{2}+Q^{2}}\delta \left( \overrightarrow{R}_{1}-\overrightarrow{R}%
_{2}\right) 
\]
In this case the equality (\ref{InvCoul}) can be written as

\begin{equation}
\left( U^{-1}\right) _{\overrightarrow{R}_{1},\overrightarrow{R}%
_{2}}\rightarrow \frac{1}{4\pi e^{2}}\left( Q^{2}-\overrightarrow{\nabla }_{%
\overrightarrow{R}_{1}}^{2}\right) \delta \left( \overrightarrow{R}_{1}-%
\overrightarrow{R}_{2}\right)  \label{InvCoulSlow}
\end{equation}
The expression (\ref{InvCoulSlow}) is correct in the case of the slow change
of the fields $\Phi $ at the scale of the order of the distance between
neighboring chains. When the characteristic distances $\mid \overrightarrow{R%
}_{1}-\overrightarrow{R}_{2}\mid $ satisfy the inequality $1/\Lambda <<\mid 
\overrightarrow{R}_{1}-\overrightarrow{R}_{2}\mid <<1/\widetilde{\Lambda }$
, the non-local character of the function $C\left( \mid \overrightarrow{R}%
_{1}-\overrightarrow{R}_{2}\mid \right) $ in (\ref{InvCoul}) is essential.
Using the equality (\ref{InvCoul}), the expression for $\left( U^{-1}\right)
_{\overrightarrow{R}_{1},\overrightarrow{R}_{2}}$ can be represented in the
form

\begin{equation}
\left( U^{-1}\right) _{\overrightarrow{R}_{1},\overrightarrow{R}_{2}}=\frac{%
\widetilde{\Lambda }^{2}}{4\pi e^{2}}\left( \frac{\Lambda ^{2}}{\pi ^{2}}%
\delta _{\overrightarrow{R}_{1},\overrightarrow{R}_{2}}-\widetilde{\Lambda }%
^{2}C\left( \delta \overrightarrow{R}\right) \right) \text{ \ \ \ }
\label{InvCR}
\end{equation}
\bigskip The general property of $\left( U^{-1}\right) _{\overrightarrow{R}%
_{1},\overrightarrow{R}_{2}}$ is the negative sign of $\left( U^{-1}\right)
_{\overrightarrow{R}_{1},\overrightarrow{R}_{2}}$ for $\overrightarrow{R}%
_{1}\neq \overrightarrow{R}_{2}$ under the condition $Q\lambda <<1$.

3. Green functions $G_{\phi }^{\left( \pm \right) }$ can be calculated
exactly [8] 
\begin{equation}
G_{\phi }^{\left( \pm \right) }\left( x,y,\overrightarrow{R}\right) =\exp
\left( \theta _{\overrightarrow{R}}^{\left( \pm \right) }\left( x\right)
-\theta _{\overrightarrow{R}}^{\left( \pm \right) }\left( y\right) \right)
G_{0}^{\left( \pm \right) }\left( x-y,\overrightarrow{R}\right) ,  \label{8}
\end{equation}
where $x=\left( t,z\right) $ and the Fourier transformations of $\theta
^{\left( \pm \right) }\left( x\right) $ are connected with the plasmon field 
$\phi $ by the equation

\begin{equation}
\theta _{\overrightarrow{R}}^{\left( \pm \right) }\left( k,\omega \right) =-i%
\frac{i\omega \pm v_{F}k}{\omega ^{2}+\left( v_{F}k\right) ^{2}}\phi _{%
\overrightarrow{R}}\left( k,\omega \right)  \label{9}
\end{equation}
In the Eq. (\ref{8}) the functions $G_{0}^{\left( \pm \right) }\left( x-y,%
\overrightarrow{R}\right) $ are the Green functions of the non-interacting
quasi-particles located at the chain $\overrightarrow{R}$.

\[
G_{0}^{\left( \pm \right) }\left( \omega ,p_{z},\overrightarrow{R}\right) =%
\frac{1}{i\omega \mp v_{F}\left( p_{z}\mp p_{F}\right) } 
\]
Our goal is the calculation of the effective action for the plasmon field $%
\Phi $. To obtain the effective action we consider the generation functional
Z

\begin{equation}
Z=\int D\psi D\overline{\psi }D\phi D\Phi ^{+}D\Phi \exp \left\{
S_{L}+S_{sources}\right\} ,  \label{GenFunc}
\end{equation}
where $S_{sources}$ depends on the sources $J,\overline{J}$\ \ and has the
form

\[
S_{sources}=i\sum\limits_{\overrightarrow{R}}\int d^{2}x\left[ \Phi
^{+}\left( x,\overrightarrow{R}\right) J\left( x,\overrightarrow{R}\right) +%
\overline{J}\left( x,\overrightarrow{R}\right) \Phi \left( x,\overrightarrow{%
R}\right) \right] 
\]
The integration over the Fermi fields in the expression for the generation
potential Eq. (\ref{GenFunc}) with the use of the Eq. (\ref{2}) gives

\begin{equation}
Z=\int D\psi D\overline{\psi }D\phi D\Phi ^{+}D\Phi \exp \left\{
S_{Pl}+S_{sources}\right\} ,  \label{GenFuncPl}
\end{equation}
where

\begin{equation}
S_{Pl}=S_{Det}\left[ \phi ,\Phi \right] +S_{forward}^{\left( 0\right) }\left[
\phi \right] +S_{back}^{\left( 0\right) }\left[ \Phi \right] +S_{sources}
\label{12}
\end{equation}

\begin{equation}
S_{Det}\left[ \phi ,\Phi \right] =SpLn\left[ \left( 
\begin{array}{cc}
\left( G_{\phi }^{\left( +\right) }\right) ^{-1} & -i\Phi \\ 
-i\Phi ^{+} & \left( G_{\phi }^{\left( -\right) }\right) ^{-1}
\end{array}
\right) \right]  \label{Det0}
\end{equation}
Note that for $\Phi =0$ the part of the action $S_{Det}\left[ \phi ,\Phi 
\right] $ can be calculated exactly [8]. As a result, the action $S_{Pl}$
can be transformed to the form

\begin{equation}
S_{Pl}=S_{Det}\left[ \Delta \right] +S_{forward}\left[ \theta \right]
+S^{int}\left[ \theta ,\Delta \right] +S_{sources},  \label{SPlasmon}
\end{equation}
where

\begin{equation}
S_{Det}\left[ \Delta \right] =SpLn\left[ \left( 
\begin{array}{cc}
\left( G_{0}^{\left( +\right) }\right) ^{-1} & -i\Delta \\ 
-i\Delta ^{+} & \left( G_{0}^{\left( -\right) }\right) ^{-1}
\end{array}
\right) \right]  \label{DetDelta}
\end{equation}

\begin{equation}
S_{forward}\left[ \theta \right] =-\int d\omega dk_{z}d^{2}\vec{k}_{\perp }%
\left[ \theta (-\omega ,-k_{z},-\vec{k}_{\perp })D_{0}^{-1}\left( \omega
,k_{z},\vec{k}_{\perp }\right) \theta (\omega ,k_{z},\vec{k}_{\perp })\right]
\label{ForwardScatt}
\end{equation}

\begin{equation}
S^{int}\left[ \theta ,\Delta \right] =-\int d^{2}x\sum\limits_{%
\overrightarrow{R},\overrightarrow{R}^{\prime }}e^{-\theta \left( x,%
\overrightarrow{R}\right) }\Delta ^{+}\left( x,\overrightarrow{R}\right)
U^{-1}\left( \overrightarrow{R},\overrightarrow{R}^{\prime }\right) \Delta
\left( x,\overrightarrow{R}^{\prime }\right) e^{\theta \left( x,%
\overrightarrow{R}^{\prime }\right) }  \label{SIntThetaDelta}
\end{equation}
In these expressions the new fields $\Delta $ , $\Delta ^{+}$ are introduced
instead of the fields $\Phi $ , $\Phi ^{+}$. They obey the equalities

\begin{equation}
\Delta \left( x,\overrightarrow{R}\right) =\Phi \left( x,\overrightarrow{R}%
\right) e^{-\theta \left( x,\overrightarrow{R}\right) }\text{ \ ; \ \ }%
\Delta ^{+}\left( x,\overrightarrow{R}\right) =\Phi ^{+}\left( x,%
\overrightarrow{R}\right) e^{\theta \left( x,\overrightarrow{R}\right) }
\label{DeltaField}
\end{equation}
The field $\theta $ is connected with the fields $\theta ^{\left( \pm
\right) }$ by the following way

\begin{equation}
\theta \left( k,\omega ,\overrightarrow{R}\right) =\theta ^{\left( +\right)
}\left( k,\omega ,\overrightarrow{R}\right) -\theta ^{\left( -\right)
}\left( k,\omega ,\overrightarrow{R}\right) =-2i\frac{v_{F}k}{\omega
^{2}+\left( v_{F}k\right) ^{2}}\phi \left( k,\omega ,\overrightarrow{R}%
\right)  \label{Theta}
\end{equation}
The inverse propagator of the $\theta $-field $D_{0}^{-1}\left( \omega
,k_{z},\vec{k}_{\perp }\right) $ has the form

\begin{equation}
D_{0}^{-1}\left( \omega ,k_{z},\vec{k}_{\perp }\right) =\frac{\left( \omega
^{2}+\left( v_{F}k_{z}\right) ^{2}\right) ^{2}}{8\left( v_{F}k_{z}\right)
^{2}}\left[ V^{-1}\left( k_{z},\vec{k}_{\perp }\right) -\Pi _{0}^{\left(
f\right) }\left( \omega ,k_{z}\right) \right]  \label{DProp}
\end{equation}
where $\Pi _{0}^{\left( f\right) }\left( \omega ,k_{z},\overrightarrow{k}%
_{\perp }\right) $ is the zero polarization operator with the small momentum
transfer $k_{z}$ ($k_{z}<<Q$)

\begin{equation}
\Pi _{0}^{\left( f\right) }\left( \omega ,k_{z},\overrightarrow{k}_{\perp
}\right) =-\frac{\kappa ^{2}}{4\pi e^{2}}\frac{\left( v_{F}k_{z}\right) ^{2}%
}{\omega ^{2}+\left( v_{F}k_{z}\right) ^{2}},  \label{PolOp}
\end{equation}
where $\kappa ^{2}$\ is the squared screening momentum ( $\kappa
^{2}=4e^{2}/\lambda ^{2}v_{F}=4\alpha \Lambda ^{2}/\pi ^{2}$).

Note that the field $\theta \left( t,z,\overrightarrow{R}\right) $ has only
the real values. This statement is a simple consequence of the expression (%
\ref{Theta}). The expressions (\ref{DetDelta}) and (\ref{SIntThetaDelta})
can be obtained, for example, from the consideration of the diagrams which
are the terms of a series expansion of (\ref{Det0}) in powers of the plasmon
field $\Phi $.

The simple expansion of $S_{Det}\left[ \Delta \right] $ in powers of the
field $\Delta $\ \ up to second power alone gives

\begin{equation}
S_{Det}\left[ \Delta \right] \thickapprox \Delta ^{+}\Pi _{Q}\Delta
\label{SDetApp}
\end{equation}
where $\Pi _{Q}=-\left( \Lambda ^{2}/2\pi ^{3}v_{F}\right) \ln \left(
\epsilon _{F}/\mid \Delta \mid \right) =-\left( \kappa ^{2}/8\pi
e^{2}\right) \ln \left( \epsilon _{F}/\mid \Delta \mid \right) $.

4. First of all, we consider the approximation for which the dependence of $%
V_{core}\left( \vec{k}_{\perp }\right) $ and $U_{core}\left( \vec{k}_{\perp
}\right) $\ ((\ref{VPseudoPot}), (\ref{UPseudoPot})) on $\vec{k}_{\perp }$
is neglected. This dependence will be considered below in the part 5 of this
paper. In this approximation the values $V_{core}$ and $U_{core}$\ can be
considered as equal values (due to the condition $Q\lambda <<1$ ). This
allows us in the case of the spinless liquid to neglect $V_{core}$ and $%
U_{core}$ and to take into account the long-range part of the Coulomb
interaction only (first terms in the sums (\ref{VPseudoPot}), (\ref
{UPseudoPot})). This approximation is absolutely correct in the case of
small momentums $k_{\perp }\lambda <<1$. In this part of the paper we
suppose that this approximation can be used for $k_{\perp }\lambda \thicksim
1$ too. Thus the action Eq. (\ref{SPlasmon}) can be represented in the form

\begin{equation}
S_{Pl}^{\left( slow\right) }\left[ \theta ,\Delta \right] =S_{Det}\left[
\Delta \right] +S_{forward}^{(slow)}\left[ \theta \right] +S_{back}^{(slow)}%
\left[ \Delta \right] +S_{int}^{(slow)}\left[ \theta ,\Delta \right]
\label{SSlow}
\end{equation}
where $S_{Det}\left[ \Delta \right] $ has the form Eq. (\ref{DetDelta}). The
terms $S_{forward}^{(slow)}\left[ \theta \right] $ and $S_{back}^{(slow)}%
\left[ \Delta \right] $ have the forms (\ref{SForward0}) and (\ref{SBack0}),
correspondingly, in which $V^{-1}\left( k_{z},\vec{k}_{\perp }\right) $ and $%
U^{-1}\left( \vec{k}_{\perp }\right) $ should be substituted by $%
V^{-1}\left( k_{z},\vec{k}_{\perp }\right) =\left( k_{z}^{2}+\vec{k}_{\perp
}^{2}\right) /4\pi e^{2}$ and $U^{-1}\left( \vec{k}_{\perp }\right) =\left(
Q^{2}+\vec{k}_{\perp }^{2}\right) /4\pi e^{2}$. Besides, the field $\Delta $
has to be substituted instead of $\Phi $.

In this case the part of the plasmon action $S_{int}^{\left( slow\right) }%
\left[ \theta ,\Delta \right] $ , which describes the interaction between
the fields $\theta $ and $\Delta $, can be represented as a sum of two terms

\begin{equation}
S_{int}^{(slow)}\left[ \theta ,\Delta \right] =S_{1}^{int}\left[ \theta
,\Delta \right] +S_{2}^{int}\left[ \theta ,\Delta \right] ,  \label{SSlInt}
\end{equation}
where

\begin{equation}
S_{1}^{int}\left[ \theta ,\Delta \right] =\frac{\lambda ^{2}}{4\pi e^{2}}%
\sum\limits_{\overrightarrow{R}}\int dtdz\left[ \Delta ^{+}\Delta \left( 
\overrightarrow{\nabla }_{\perp }\theta \right) ^{2}\right]
\label{SSlowInt1}
\end{equation}

\begin{equation}
S_{2}^{int}\left[ \theta ,\Delta \right] =i\frac{\lambda ^{2}}{4\pi e^{2}}%
\sum\limits_{\overrightarrow{R}}\int dtdzd^{2}\overrightarrow{R}\left( 
\overrightarrow{j}\left( \overrightarrow{\nabla }_{\perp }\theta \right)
\right)  \label{SSlowInt2}
\end{equation}

\[
\overrightarrow{j}=-i\left( \Delta ^{+}\left( \overrightarrow{\nabla }%
_{\perp }\Delta \right) -\Delta \left( \overrightarrow{\nabla }_{\perp
}\Delta ^{+}\right) \right) 
\]
Note, that the presence of the term $S_{1}^{int}\left[ \theta ,\Delta \right]
$ leads to the divergency of the functional integral for the large values of 
$\overrightarrow{\nabla }_{\perp }\theta $ and, for this reason, this means
the necessity of the consideration of the large momentum $\overrightarrow{k}%
_{\perp }$ contribution or the lattice formulation of the plasmon action (%
\ref{SPlasmon}). The lattice formulation will be given below in part 5.

Thus the effective action for the plasmon fields can be represented as

\begin{eqnarray*}
&&S_{Pl}^{\left( slow\right) }[\Delta ,\theta ]=S_{Det}\left[ \Delta \right]
-\int \theta D_{0}^{-1}\theta \\
&&-\frac{\lambda ^{2}}{4\pi e^{2}}\sum\limits_{\overrightarrow{R}}\int dtdz%
\left[ \Delta ^{+}\left( Q^{2}-\overrightarrow{\nabla }_{\perp }^{2}-\left( 
\overrightarrow{\nabla }_{\perp }\theta \right) ^{2}\right) \Delta \right] \\
&&+\frac{\lambda ^{2}}{4\pi e^{2}}\sum\limits_{\overrightarrow{R}}\int dtdz%
\left[ \left( \Delta ^{+}\left( \overrightarrow{\nabla }_{\perp }\Delta
\right) -\Delta \left( \overrightarrow{\nabla }_{\perp }\Delta ^{+}\right)
\right) \left( \overrightarrow{\nabla }_{\perp }\theta \right) \right]
\end{eqnarray*}
Here $S_{Det}\left[ \Delta \right] $ is determined by the equality (\ref
{DetDelta}).\ It is convenient to introduce the amplitude and the phase of
the field $\Delta $ in the following way

\begin{eqnarray}
\Delta &=&\rho e^{i\chi }  \label{Delt} \\
\Delta ^{+} &=&\rho e^{-i\chi }  \nonumber
\end{eqnarray}
Thus we obtain

\begin{equation}
S_{slow}=SpLn\left[ \widehat{G}^{-1}\left( \Delta \right) \right] -\sum \int
\theta D_{0}^{-1}\theta -\frac{\lambda ^{2}}{4\pi e^{2}}\sum\limits_{%
\overrightarrow{R}}\int \left( Q^{2}\rho ^{2}+\left( \overrightarrow{\nabla }%
_{\perp }\rho \right) ^{2}-\rho ^{2}\left( \overrightarrow{\nabla }_{\perp
}\left( \theta +i\chi \right) \right) ^{2}\right)  \label{S1}
\end{equation}
where $D_{0}^{-1}$ is determined by the equality (\ref{DProp}) and $S_{Det}%
\left[ \Delta \right] =SpLn\left[ \widehat{G}^{-1}\left( \Delta \right) %
\right] $ (Eq. (\ref{DetDelta})).

We introduce a new variable $\eta $ instead of $\theta $

\begin{equation}
\eta =\theta +i\chi  \label{Etta}
\end{equation}

After that we have

\begin{eqnarray}
S_{slow} &=&SpLn\left[ \widehat{G}^{-1}\left( \Delta \right) \right] -\sum
\int \eta \left( D_{0}^{-1}-\frac{1}{4\pi e^{2}}\overleftarrow{\nabla }%
_{\perp }\rho ^{2}\overrightarrow{\nabla }_{\perp }\right) \eta +2i\sum \int
\eta D_{0}^{-1}\chi  \label{S2} \\
&&+\sum \int \chi \left( D_{0}^{-1}\right) \chi -\frac{\lambda ^{2}}{4\pi
e^{2}}\sum\limits_{\overrightarrow{R}}\int \left( Q^{2}\rho ^{2}+\left( 
\overrightarrow{\nabla }_{\perp }\rho \right) ^{2}\right)  \nonumber
\end{eqnarray}
where the operator $\overrightarrow{\nabla }_{\perp }$ acts on the fields
which are situated at the right side of this operator and $\overleftarrow{%
\nabla }_{\perp }$ acts on the fields which are situated at the left side of
this operator.\ 

The integration over $\eta $ gives

\begin{eqnarray}
S_{slow}\left[ \Delta \right] &=&SpLn\left[ \widehat{G}^{-1}\left( \Delta
\right) \right] -\frac{1}{2}SpLn\left[ D^{-1}\left( \rho \right) \right] -%
\frac{\lambda ^{2}}{4\pi e^{2}}\sum\limits_{\overrightarrow{R}}\int \left(
Q^{2}\rho ^{2}+\left( \overrightarrow{\nabla }_{\perp }\rho \right)
^{2}\right)  \label{SEff1} \\
&&-\int \chi D_{0}^{-1}\left( D-D_{0}\right) D_{0}^{-1}\chi  \nonumber
\end{eqnarray}
where

\begin{equation}
D^{-1}=D_{0}^{-1}-\frac{1}{4\pi e^{2}}\overleftarrow{\nabla }_{\perp }\rho
^{2}\overrightarrow{\nabla }_{\perp }  \label{D}
\end{equation}
The last term in (\ref{SEff1}) in the approximation of the small value of $%
\rho ^{2}$ can be represented in the form

\[
\int \chi D_{0}^{-1}\left( D-D_{0}\right) D_{0}^{-1}\chi =\int \chi
D_{0}^{-1}\frac{D_{0}^{-1}-D^{-1}}{D_{0}^{-1}D^{-1}}D_{0}^{-1}\chi =\frac{1}{%
4\pi e^{2}}\int \rho ^{2}\left( \overrightarrow{\nabla }_{\perp }\chi
\right) ^{2} 
\]
After that we obtain the following expression for the effective plasmon
action

\begin{eqnarray}
S_{slow}\left[ \Delta \right] &=&SpLn\left[ \widehat{G}^{-1}\left( \Delta
\right) \right] -\frac{1}{2}SpLn\left[ D^{-1}\left( \rho \right) \right] -%
\frac{\lambda ^{2}}{4\pi e^{2}}\sum\limits_{\overrightarrow{R}}\int
dtdz\left( Q^{2}\rho ^{2}+\left( \overrightarrow{\nabla }_{\perp }\rho
\right) ^{2}\right)  \label{SEffRoHi} \\
&&-\frac{\lambda ^{2}}{4\pi e^{2}}\sum\limits_{\overrightarrow{R}}\int
dtdz\left( \rho ^{2}\left( \overrightarrow{\nabla }_{\perp }\chi \right)
^{2}\right)  \nonumber
\end{eqnarray}

There is nonzero saddle-point field for this action.\ The saddle-point
equations in the case of the constant $\rho $\ as a function of $t$, z and $%
\overrightarrow{R}$ coordinates \ and for the field $\chi $\ have the form 
\begin{equation}
-\frac{1}{v_{F}}\int \frac{d\omega d\xi d^{2}k_{\perp }}{\left( 2\pi \right)
^{4}}\frac{1}{\omega ^{2}+\xi ^{2}-\rho ^{2}}+\frac{1}{2v_{F}}\int \frac{%
d\omega d\xi d^{2}k_{\perp }}{\left( 2\pi \right) ^{4}}\frac{V^{-1}}{%
D_{0}^{-1}-\rho ^{2}V^{-1}}-\frac{Q^{2}}{4\pi e^{2}}-\frac{\left( 
\overrightarrow{\nabla }_{\perp }\chi \right) ^{2}}{4\pi e^{2}}=0
\end{equation}

\begin{equation}
\overrightarrow{\nabla }_{\perp }^{2}\chi =0  \label{SaddleEq2}
\end{equation}
The possible solution of these saddle-point equations can be represented as 
\[
\rho =i\rho _{0}\ 
\]
where $\rho _{0}$ is the constant real number field and the solution for the
phase $\chi $ is 
\begin{equation}
\chi =\overrightarrow{K}_{\perp }\overrightarrow{R}  \label{SaddleHi}
\end{equation}
Here $\overrightarrow{K}_{\perp }$ is an arbitrary vector lying in the cell
of the reciprocal lattice. The value of $\rho _{0}$ has the form

\begin{equation}
\rho _{0}=\epsilon _{F}\exp \left( -\frac{\lambda ^{2}\left( Q^{2}+%
\overrightarrow{K}_{\perp }^{2}\right) }{2\alpha }\right)  \label{SaddleRo}
\end{equation}
The statistical sum $Z$\ has the maximum value for $\overrightarrow{K}%
_{\perp }=\left( \pi /\lambda ,\pi /\lambda \right) $. Thus the ground state
of the system is inhomogeneous with a charge-density wave which has the wave
vector along the direction of chains equal to $2p_{F}$\ and in the
perpendicular direction equal to $\overrightarrow{K}_{\perp }=\left( \pi
/\lambda ,\pi /\lambda \right) $.

The fluctuations $\delta \rho $\ of the field $\rho $ can be taken into
account by the following representation of the field $\rho $ 
\[
\rho =i\rho _{0}\ +\delta \rho 
\]

Note that in the approximation of the slow variation of the fields $\rho $
and $\chi \ $ in the time $t$ and z-coordinate the calculation of the first
two terms in (\ref{SEffRoHi}) gives

\[
SpLn\left[ \widehat{G}^{-1}\left( \Delta \right) \right] =\frac{1}{4\pi v_{F}%
}\sum\limits_{\overrightarrow{R}}\int dtdz\left[ \rho _{0}^{2}\ln \left( 
\frac{\epsilon _{F}^{2}}{\rho _{0}^{2}}\right) \right] 
\]

\[
SpLn\left[ D^{-1}\left( \rho \right) \right] =\frac{2}{4\pi v_{F}}%
\sum\limits_{\overrightarrow{R}}\int dtdz\left[ \rho _{0}^{2}\ln \left( 
\frac{\epsilon _{F}^{2}}{\rho _{0}^{2}}\right) \right] 
\]

5. In this part of the paper the dependence of $V_{core}\left( \vec{k}%
_{\perp }\right) $ and $U_{core}\left( \vec{k}_{\perp }\right) $ on $\vec{k}%
_{\perp }$ are taken into account. This is essentially for involving of the
short-range fluctuations. If there is an instability of the system in the
plasmon channel with the momentum transfer $Q=2p_{F}$ , the plasmon
effective action, obtained after the integration over the field $\theta $
and dependent on the field $\Delta $ alone,$\ $tends to zero near the
instability point. Thus, near the instability point the exponent in the
expression (\ref{GenFuncPl}) can be expanded into a series in action $S_{Det}%
\left[ \Delta \right] +S^{int}\left[ \theta ,\Delta \right] $, and after
that the integral over the field $\theta $ can be calculated due to the
Gaussian form of this integral. \ In this case the contribution of $S^{int}%
\left[ \theta ,\Delta \right] $ can be represented in the form

\begin{equation}
<S^{int}\left[ \theta ,\Delta \right] >_{\theta }=-\int d^{2}x\sum\limits_{%
\overrightarrow{R},\overrightarrow{R}^{\prime }}\Delta ^{+}\left( x,%
\overrightarrow{R}\right) U^{-1}\left( \overrightarrow{R},\overrightarrow{R}%
^{\prime }\right) \Delta \left( x,\overrightarrow{R}^{\prime }\right)
C_{\theta }\left( \delta \overrightarrow{R}\right)  \label{SIntAv}
\end{equation}
where

\[
C_{\theta }\left( \delta \overrightarrow{R}\right) =<e^{\theta \left( x,%
\overrightarrow{R}^{\prime }\right) -\theta \left( x,\overrightarrow{R}%
\right) }>_{\theta } 
\]
Here $\delta \overrightarrow{R}=\overrightarrow{R}^{\prime }-$ $%
\overrightarrow{R}$,

\begin{equation}
<e^{\theta \left( x,\overrightarrow{R}^{\prime }\right) -\theta \left( x,%
\overrightarrow{R}\right) }>_{\theta }=\int D\theta \exp \left\{ \theta
\left( x,\overrightarrow{R}^{\prime }\right) -\theta \left( x,%
\overrightarrow{R}\right) +S_{forward}\left[ \theta \right] \right\}
\label{ThetaCorr}
\end{equation}
The integration over $\theta $ in (\ref{ThetaCorr}) yields \ 

\begin{equation}
<e^{\theta \left( t,z,\overrightarrow{R}^{\prime }\right) -\theta \left( t,z,%
\overrightarrow{R}\right) }>_{\theta }=\exp \left\{ \frac{1}{2}\int \frac{%
d\omega dk_{z}d^{2}\vec{k}_{\perp }}{\left( 2\pi \right) ^{4}}\left( 1-\cos
\left( \overrightarrow{k}_{\perp }\delta \overrightarrow{R}\right) \right)
D(\omega ,k_{z},\vec{k}_{\perp })\right\}  \label{ThetaCorr1}
\end{equation}
The propagator $D(\omega ,k_{z},\vec{k}_{\perp })$ obeys the Eq. (\ref{DProp}%
). The calculation of the integral (\ref{ThetaCorr1}) for $C_{\theta }\left(
\delta \overrightarrow{R}\right) $\ gives

\begin{equation}
C_{\theta }\left( \delta \overrightarrow{R}\right) =<e^{\theta \left( t,z,%
\overrightarrow{R}^{\prime }\right) -\theta \left( t,z,\overrightarrow{R}%
\right) }>_{\theta }=\exp \left[ f\left( \delta \overrightarrow{R}\right) %
\right]  \label{C}
\end{equation}
The function $f\left( \delta \overrightarrow{R}\right) $\ has the form

\begin{eqnarray}
f\left( \delta \overrightarrow{R}\right) &=&\gamma \alpha \xi \left[ \ln
\left( \frac{\Lambda _{\infty }}{\kappa }\right) -K_{0}\left( \kappa \mid
\delta \overrightarrow{R}\mid \right) \right] \text{ \ \ \ \ \ \ \ for \ \ }%
\mid \delta \overrightarrow{R}\mid >>1/\Lambda _{\infty }\   \label{f} \\
f\left( \delta \overrightarrow{R}\right) &=&0\text{ \ \ \ \ \ \ \ \ \ \ \ \
\ \ \ \ \ \ \ \ \ \ \ \ \ \ \ \ \ \ \ \ \ \ \ \ \ \ \ \ \ \ \ \ \ \ \ \ \ \
\ \ \ \ \ \ \ \ \ \ \ \ \ \ \ \ \ for \ \ \ }\delta \overrightarrow{R}=0 
\nonumber
\end{eqnarray}

\ $\gamma =2/\pi $; $\ $ $\xi =\ln \left( \varepsilon _{F}/\mid \Delta \mid
\right) $\ ; \ \ $\Lambda _{\infty }=1/a$.

Thus, the part of the effective action $<S^{int}\left[ \theta ,\Delta \right]
>_{\theta }$ leads to the renormalization of $U^{-1}\left( \overrightarrow{R}%
,\overrightarrow{R}^{\prime }\right) $ in (\ref{SPlasmon}) and the
renormalized quantity $U_{eff}^{-1}\left( \overrightarrow{R},\overrightarrow{%
R}^{\prime }\right) $ \ has the form

\begin{equation}
U_{eff}^{-1}\left( \delta \overrightarrow{R}\right) =U^{-1}\left( \delta 
\overrightarrow{R}\right) C_{\theta }\left( \delta \overrightarrow{R}\right)
\label{InvUeffCoord}
\end{equation}
Thus\bigskip

\begin{equation}
U_{eff}^{-1}\left( \delta \overrightarrow{R}\right) =\frac{1}{4\pi e^{2}\pi
^{2}}\widetilde{\Lambda }^{2}\Lambda ^{2}\delta _{\overrightarrow{R},%
\overrightarrow{R}^{\prime }}-\frac{\widetilde{\Lambda }^{4}}{4\pi
e^{2}\left( 2\pi \right) }K_{0}\left( \mid \delta \overrightarrow{R}\mid 
\widetilde{\Lambda }_{Q}\right) \exp \left[ f\left( \delta \overrightarrow{R}%
\right) \right]  \label{InvUeffR}
\end{equation}
The calculation of the Fourier component of $U_{eff}^{-1}$ \ gives 
\begin{equation}
U_{eff}^{-1}\left( \overrightarrow{k_{\perp }}\right) =\frac{1}{4\pi e^{2}}%
\widetilde{\Lambda }^{2}\left[ 1-A\left( \overrightarrow{k_{\perp }},\xi
\right) \left( \frac{\Lambda _{\infty }}{\widetilde{\Lambda }_{Q}}\right)
^{\gamma \alpha \xi }\left( \frac{\widetilde{\Lambda }}{\widetilde{\Lambda }%
_{Q}}\right) ^{2}\right]  \label{InvUeffMom}
\end{equation}
where

\begin{equation}
A\left( \overrightarrow{k_{\perp }},\xi \right) =\int\limits_{0}^{\Lambda /%
\widetilde{\Lambda }_{Q}}dxJ_{0}\left( x\mid \overrightarrow{k_{\perp }}\mid
/\widetilde{\Lambda }_{Q}\right) K_{0}\left( x\right) x^{\gamma \alpha \xi
+1}  \label{A1}
\end{equation}
The momentum $\widetilde{\Lambda }_{Q}$ is $\widetilde{\Lambda }_{Q}=\sqrt{%
\widetilde{\Lambda }^{2}+Q^{2}}$. Assuming $\Lambda /\widetilde{\Lambda }%
_{Q}<<1$, the above limit of the integral (\ref{A1}) can be replaced by the
infinity. Note that $U_{eff}^{-1}$ decreases with the increase of the value
of $\mid \overrightarrow{k_{\perp }}\mid $ for the small $\mid 
\overrightarrow{k_{\perp }}\mid $ \ $\left( \mid \overrightarrow{k_{\perp }}%
\mid /\widetilde{\Lambda }_{Q}<<1\right) $. The definition of $A\left( 
\overrightarrow{k_{\perp }},\xi \right) $ (\ref{A1}) gives $A\left( 
\overrightarrow{k_{\perp }}=0,\xi =0\right) $\ $=1$. Thus in the
approximation $\alpha \xi <<1$ and $\mid \overrightarrow{k_{\perp }}\mid <<1/%
\widetilde{\Lambda }$ we can put $A\left( \overrightarrow{k_{\perp }},\xi
\right) $ equal to unity $A\left( \overrightarrow{k_{\perp }},\xi \right) =1 
$ in (\ref{InvUeffMom}).

The effective action of the plasmon field $\Delta $\ can be represented in
the form \bigskip

\begin{equation}
S_{eff}=SpLn\left[ \widehat{G}^{-1}\left( \Delta \right) \right] +<S^{int}%
\left[ \theta ,\Delta \right] >_{\theta }  \label{EffAction}
\end{equation}

\[
<S^{int}\left[ \theta ,\Delta \right] >_{\theta }=-\int d^{2}x\int \frac{%
d^{2}\overrightarrow{k_{\perp }}}{\left( 2\pi \right) ^{2}}\Delta ^{+}\left(
x,\overrightarrow{k_{\perp }}\right) U_{eff}^{-1}\left( \overrightarrow{%
k_{\perp }}\right) \Delta \left( x,\overrightarrow{k_{\perp }}\right) 
\]
Expanding $SpLn\left[ \widehat{G}^{-1}\left( \Delta \right) \right] $\ in
the effective action $S_{eff}$ (\ref{EffAction})\ in field $\Delta $, we
obtain\ the quadratic action for the $\Delta $ field and the inverse
propagator of this field has the form

\begin{equation}
\Gamma ^{-1}\left( \overrightarrow{k_{\perp }},\xi \right) =-\Pi
_{Q}+U_{eff}^{-1}\left( \overrightarrow{k_{\perp }},\xi \right)
\label{Gamma}
\end{equation}
Here $\Pi _{Q}=\left( -\kappa ^{2}/8\pi e^{2}\right) \xi $ where $\kappa
^{2}=4\alpha \Lambda ^{2}/\pi ^{2}$ and $\xi $ is the logarithmic variable $%
\ \xi =\ln \left( \epsilon _{F}/\mid \omega \mid \right) $ .

Now we will consider the pole of the propagator $\Gamma $ when $%
\overrightarrow{k}_{\perp }=0$. Note that the situation with $%
\overrightarrow{k_{\perp }}\neq 0$\ is analogous to that which has been
considered in the part 4. For $\overrightarrow{k_{\perp }}=0$\ \ the
equation determining the pole $\xi _{0}=\ln \left( \epsilon _{F}/\mid \omega
_{0}\mid \right) $ of $\Gamma \left( \overrightarrow{k}_{\perp }=0,\text{ }%
\xi \right) $ takes the form

\begin{equation}
-4\pi e^{2}\Pi _{Q}+\widetilde{\Lambda }^{2}\left[ 1-\left( \frac{\Lambda
_{\infty }}{\widetilde{\Lambda }_{Q}}\right) ^{\gamma \alpha \xi }\left( 
\frac{\widetilde{\Lambda }}{\widetilde{\Lambda }_{Q}}\right) ^{2}\right] =0
\label{GapEq}
\end{equation}

Using the expression for $\Pi _{Q}$ and $\widetilde{\Lambda }^{2}=(2/\pi
)\Lambda ^{2}\ln ^{-1}\left( \frac{\Lambda _{\infty }}{\Lambda }\right) $ \
the Eq. (\ref{GapEq}) can be represented in the form

\begin{equation}
\frac{1}{2}\gamma \alpha \xi \ln \left( \frac{\Lambda _{\infty }}{\Lambda }%
\right) +\left[ 1-\left( \frac{\widetilde{\Lambda }}{\widetilde{\Lambda }_{Q}%
}\right) ^{2}\exp \left( \gamma \alpha \xi \ln \left( \frac{\Lambda _{\infty
}}{\widetilde{\Lambda }_{Q}}\right) \right) \right] =0  \label{GapEq1}
\end{equation}

It can readily be seen that the right-hand side of the equation (\ref{GapEq}%
) is positive for $\xi =0$. However, as $\xi $ increases, the right-hand
side of (\ref{GapEq}) \ becomes negative. Thus, the solution of this
equation exists and for the supposition $Q<<\widetilde{\Lambda }$ has the
following value within the logarithmic accuracy

\begin{equation}
\mid \omega _{0}\mid \thicksim \epsilon _{F}\exp \left( -\frac{2Q^{2}}{%
\widetilde{\Lambda }^{2}\gamma \alpha \ln \left( \frac{\Lambda _{\infty }}{%
\Lambda }\right) }\right) =\epsilon _{F}\exp \left( -\frac{\lambda ^{2}Q^{2}%
}{2\alpha }\right)  \label{Gap}
\end{equation}
Note, that the Eq. (\ref{Gap}) is correct when $\frac{\lambda ^{2}Q^{2}}{%
2\alpha }\thicksim Q^{2}/\kappa ^{2}<<1$. The solution (\ref{Gap}) coincides
with the solution (\ref{SaddleRo}) for $\overrightarrow{K}_{\perp }=0$. Here 
$\omega _{0}$ corresponds to the pole of $\Gamma \left( \overrightarrow{%
k_{\perp }}=0,\xi \right) $. The existence of this pole means the
instability of the system. As a result of this instability there is a CDW in
the ground state of the system. The z-component of the wave vector of this
CDW equals $2p_{F}$ . As it has been discussed in part 4 the wave vector of
this CDW has the component perpendicular to the direction of the chains and
equal to $\overrightarrow{K}_{\perp }=\left( \pi /\lambda ,\pi /\lambda
\right) $. \ The existence of the CDW results in the existence of the
dielectric gap at the Fermi surface. The value of the gap is equal to $%
\Delta _{g}=\rho _{0}$, where $\rho _{0}$ is defined by (\ref{SaddleRo}).


\begin{references}
\bibitem{BBK}  [1] K. Behnia, L. Balicas, W. Kang, D. Jerom, P. Carretta, Y.
Fogot-Revurat, C. Berthier, M. \ \ 

\ \ \ Horvatic, P. Segransan, L. Hubert, and Bourbonannais, Phys. Rev. Lett.
74, 5272 (1995).

\bibitem{ChCh}  [2] E. I. Chashechkina and P. M. Chaikin, Phys. Rev. Lett.
80, 2181 (1998).

\bibitem{Gor'kov}  [3] L. P. Gor'kov, I. E. Dzyaloshinskii, Zh. Eksp. Teor.
Fiz. {\bf 67}, 397, (1974).

\bibitem{Dzyaloshinskii}  [4] I. E. Dzyaloshinskii and A. I. Larkin, Zh.
Eksp. Teor. Fiz. {\bf 65}, 411, (1973). \ 

\bibitem{EggerGogolin}  [5] R. Egger and A. O. Gogolin, Phis. Rev. Letters 
{\bf 79}, 5082, (1997).

\bibitem{Yakovenko}  [6] V.M. Yakovenko, Phys. Rev. B {\bf 47}, 8851,
(1993)\ \ \ \ \ \ \ \ \ \ \ \ \ \ \ \ \ \ \ \ 

\bibitem{BabKoz}  [7] V. S. Babichenko, A. N. Kozlov, Pis'ma Zh. Eksp. Teor.
Fiz. {\bf 61}, (1995), 722. \ \ \ \ \ \ \ 

\bibitem{Schwinger}  [8] J. Schwinger, Phys. Rev. {\bf 128}, 2425, (1962).

\bibitem{BabOn}  [9] V. S. Babichenko and T. A. Onischenko, JETP Lett. {\bf %
26}, 68, (1977). \ 

\bibitem{Solyom}  [10]\ \ J. Solyom, Advances in Physics, 28, 201, (1979).\
\ \ \ \ \ \ \ \ \ \ \ \ \ \ 

\bibitem{KeldOn}  [11] L. V. Keldysh and T. A. Onishchenko, JETP Lett. {\bf %
24}, 69, (1976).
\end{references}
\end{document}